\documentclass[%
reprint,
 amsmath,amssymb,
 prl,
]{revtex4}

\usepackage{graphicx}
\usepackage{dcolumn}
\usepackage{bm}

\begin{document}

\preprint{Physics Review Letters}

\title{Acoustic Invisibility in Turbulent Fluids}

\author{Xun Huang}
 \email{huangxun@pku.edu.cn}
\author{Siyang Zhong}%
\affiliation{%
 State Key Laboratory of Turbulence and Complex Systems, College of Engineering, Peking University, Beijing, 100871, China
}
%




\date{\today}

\begin{abstract}
Acoustic invisibility of a cloaking system in turbulent fluids has been poorly understood.  Here we show that evident scattering would appear in turbulent wakes owing to the submergence of a classical cloaking device. The inherent mechanism is explained using our theoretical model, which eventually inspires us to develop an optimized cloaking approach. Both the near- and far-field scatted fields are examined using high order computational acoustic methods. The remarkably low scattering demonstrates the effectiveness of the proposed acoustic cloaking approach for turbulent fluid cases.

%


\end{abstract}

\maketitle


Acoustic invisibility, a  thought-provoking and compelling concept, is one of the most fascinating physical problems that can be resolved by recent cloaking technologies \cite{Pendry:sci06cloak, Chen:07Cloak}.
Classical  acoustic cloaking approaches  \cite{Chen:07Cloak, Cummer:07,Norris:08Cloak},
either for underwater ultrasound \cite{Farhat:08Cloak,Zhang:11Cloak} or airborne sound \cite{Popa:11Cloak,Sanchis:13Cloak},
 usually follow ingenious cloaking paradigms in electromagnetic waves  \cite{Pendry:sci06cloak,Leo:sci09cloak,Liu:sci09cloak,Vasquez:09PRL,Chen:12AM} whereas take no account of moving media. Fundamental restrictions, as a result, exist in the capability of acoustic invisibility for those classical cloaking approaches in turbulent fluids.


To demonstrate the above deficiency, we start with a representative example as shown in Fig.~1. The circular interior is concealed from the acoustic detections by an annular cloaking  device, which submerges in a hypothetically incompressible fluid with density $\rho_0$, the undisturbed free stream speed $U_\infty$ and the speed of sound $c_0$.
We expect that the external acoustic observers would be unaware of both the hidden interior and the cloaking device.
To achieve this objective, classical acoustic cloaking approaches  \cite{Chen:07Cloak,Cummer:07} take the assumption that the sound propagations are governed by the homogeneous wave equation that is coordinate invariance.  The method of transformation acoustics converts this annular cloak in physical space to a cylindrical cloak in the virtual space (Fig.~2A), where the hidden interior is reduced to a singular point that is concealed from the sound waves.
If we adopt  the classical cloaking approach, in the virtual space, the cloak's density and the speed of sound through the cloaking material respectively satisfy
\begin{equation}\label{e:design1}
\rho_T=\rho_0, c_T=c_0,
\end{equation}
\noindent to match the acoustic impedances between the cloak and the surrounding fluid; where $(\cdot)_T$ indicates the cloak's variables that are already transformed in the virtual space. The corresponding density and bulk modulus in physical space are eventually achieved by the associated coordinate transformation. Here we simply study the cloaking approaches in the virtual space because of its one-to-one connection to physical space.


Our numerical simulations suggest that the acting of the classical cloaking device could be detectable in the turbulent fluids. To illustrate this issue, we perform calculations of the flow field and the sound field in succession. The corresponding Reynolds number is 90,000 based on the outer radius of the cloak and $U_\infty$. All variables in the computational model are nondimensionalized using the respective references. Our detached eddy simulations reveal that intricate, evolving dynamics of three-dimensional (3D) coherent flow structures are exhibited in the turbulent wakes of the cloak (see Fig.~1). Here we have assumed that the material of the cloaking device is transparent to the sound waves but impervious to the surrounding fluids. To perform direct acoustic simulations, we further assume that the time scale of fluid dynamics is much longer than that of the sound waves, which enables the use of  Euler equations linearized about the instantaneous turbulent flow for the following acoustic simulations. The incident sound wave is of the form:
\begin{equation}\label{e:pinc}
p'_{inc}({x},y,z,t)=A e^{i\omega t-i k x}, k=\frac{k_0}{(1+M_\infty)},
\end{equation}
\noindent where $A$ is the amplitude of the wave, in this work $A=10^{-5}$; $\omega$ is angular frequency; and $k_0=\omega/c_0$ and $M_\infty=U_\infty/c_0$. Ray trajectories are calculated to keep track of the sound waves. For the sake of clarity, only one two-dimensional (2D) cross section map of the ray trajectories are shown in Fig.~1.
It can be seen that the sound waves are bent at will to successfully conceal the hidden interior. In contrast, an evident scattered field,  which would eventually disclose the acting of the cloaking device itself, appears in the turbulent wakes, owing to the abrupt and random changes of the velocity gradients in this turbulent flow. A different cloaking approach with the ability not only to control the internal ray trajectories but to suppress external scattered field in the turbulent fluids shall be required to resolve such fundamental limit.

Here we explain that an additional optimization of cloaking design would eventually reduce the scattered field posed by the turbulent fluids. The pivotal idea behind our approach is pursuant to the theoretical developments that reduce the sound propagation model (linearized Euler equations)
to the inhomogeneous wave equation in an otherwise ideal and stationary acoustic medium \cite{online}
\begin{equation}\label{e:wav4}
\frac{\partial^2}{\partial t^2}p'_{sca} - c_0^2\nabla^2p'_{sca} = s,
\end{equation}
\noindent where the surface $f=0$ defines the boundary of the cloak in the virtual space; $\mathrm{H}$ is the Heaviside step function, $\mathrm{H}(f)=1$ inside the fluids and $\mathrm{H}(f)=0$ within the cloak; and $s=\mathrm{H}(f) s_o + (1-\mathrm{H}(f)) s_i$ indicates the equivalent sound sources due to the outside fluids ($s_o$) and the internal cloak ($s_i$):
\begin{eqnarray}
 \nonumber  s_i (\mathbf{x}, t) &=&  (\rho_0c_0^2 - \rho_Tc_T^2)\frac{\partial}{\partial t}\left (\nabla\cdot\mathbf{u}'_{inc} \right) - \left(1 - \frac{\rho_0}{\rho_T}\right)c_0^2\nabla^2p'_{inc} \\
 &&  - \left( \frac{\partial^2p'_{inc}}{\partial t^2} - c_0^2\nabla^2p'_{inc} \right),  \label{e:wav5} \\
\nonumber  s_o (\mathbf{x}, t) &=&  \rho_0c_0^2\nabla\cdot(\mathbf{u}_0\cdot\nabla\mathbf{u}'_{inc}+\mathbf{u}'_{inc}\cdot\nabla\mathbf{u}_0) + \\
  && \nabla \cdot \left(p'_{inc}\mathbf{u}_0\cdot\nabla\mathbf{u}_0\right)-\nabla\cdot\left[\frac{\partial( p'_{inc}\mathbf{u}_0)}{\partial t}\right] - \left( \frac{\partial^2p'_{inc}}{\partial t^2} - c_0^2\nabla^2p'_{inc} \right), \label{e:wav6}
\end{eqnarray}
\noindent where $\mathbf{x}=(x, y, z)$ is the coordinates of the sources; $\mathbf{u}'_{inc}$ is the particle velocity of the incident wave; and $\mathbf{u}_0$ is the background flow field. The complete theoretical developments can be found in the supplementary material \cite{online}.

The above notion is examined by calculating $s$ throughout the entire domain.  The most striking finding in Fig.~2A is $|s_i|\ne 0$ within the classical cloaking device.  This peculiarity can be simply explained by recalling that the propagation of the incident sound wave in the free stream is governed by convected wave equation, while its propagation within the cloak is governed by classical wave equation. As a result, $s_i$ primarily depends on $\rho_T/\rho_0$.


The external sound source $s_o$ is determined for the given flow field and the incident wave. A question follows herein shall be whether a suitable $s_i$ can be achieved that cancels $s_o$ and eventually controls the scattered field. The cloaking design in the turbulent fluids, in accordance with the above model, i.e. equation~(3), is therefore deemed as an optimization problem,  which would optimize $\rho_T$ to render the desired sound field.  The classical cloaking design is therefore relaxed to $\rho_T c_T=\rho_0 c_0$ to maintain impedance matching. We then construct an objective function $F_c$ at the chosen observers (Fig. 2A) to evaluate the scattered field. Generally speaking, $F_c$ is the least-squares functional of $(p'_{inc}-p'_{sca})$, which solely depends on the $s_i(\rho_T)$ and $s_o$.  The optimal $\rho_T$ is solved by the following mathematical programming
\begin{equation}
\underset{\rho_T} {\mathrm{arg}~\mathrm{min}}~F_c( s_i(\rho_T), s_o),
\end{equation}
\noindent where $\rho_T$ can be inhomogeneous that results in a multi-dimensional optimization (Fig. 2B). It would be a matter of straight algebra to eventually transform this optimized cloaking design from the virtual space to physical space. 



Direct numerical simulations of sound propagation explicitly demonstrate that our cloaking approach enables acoustic invisibility in the turbulent fluids. In the following numerical studies, the objective function $F_c$ is constructed at $r=10$ and thereafter optimized to produce $\rho_T$. Figure 2C shows that the optimized cloaking design would not only exclude the sound waves from the interior but almost restore the ray trajectories in the turbulent wakes. More numerical details and results can be found in the  supplementary material \cite{online}.

To quantify the control performance in the turbulent fluids, we define the criterion of
\begin{equation}
 \mathrm{err} = 20\log_{10}\left(\frac{\overline{p'}}{\overline{p'_{inc}}}\right) \label{e:err},
\end{equation}
\noindent where $\overline{(\cdot)}$ denotes the root mean square value. Significant difference between the classical cloaking approach and the optimized cloaking approach
can thus be revealed by examining the values of $\mathrm{err}$, respectively. Figure~3 show that there are abrupt changes of $\mathrm{err}$, between $-15$ and $10\,$dB (at $r=10$) and between $-9$ and $5\,$dB  (at $r=100$), for the classical cloaking device at low $\theta$ angles, which correlate with the region of the turbulent wakes. In contrast, our optimized cloak maintains a much smaller $\mathrm{err}$, between $-1.6$ and $1.9\,$dB (at $r=10$) and between $\pm 0.6\,$dB (at $r=100$), across the full range of $\theta$. It should be mentioned that we simply use the method of steepest descent as the optimization solver in the iterative optimizations. The control performance could be further improved if alternative optimization algorithms, such as adaptive filters \cite{Elliott:87ieee}, are considered. In addition, for various instantaneous flow fields, the results appear to have statistically deterministic distributions, which suggest multiple scattering occurs in the turbulent fluids.


As for practical applications, our approach can be directly adopted in the conceptual designs.  The acoustic cloak must be able to change its design and implementation along with the change of the surrounding turbulent fluids, which requires flow predications in real time and dynamically tunable metamaterial. Both requirements could be fulfilled in the near future owing to fast  evolution of computer technologies and recent progress in advanced materials \cite{nano11,Chen:12AM}.

In this work, the potential effect of flow-induced self-noise is ignored to examine the effect of turbulent fluids on sound propagations alone. In summary, we elucidated the scattering problem due to the submergence of a cloak in the turbulent fluids by establishing a theoretical model, which plays a critical role in the follow-up development of our optimized cloaking approach. Such beneficial performance, particularly in the turbulent wakes, has been demonstrated.  We believe that the approach would be beneficial for a host of new material designs and engineering systems.

This work is supported by National Science Foundation of China (Grants 11172007 and 11322222).  We are grateful to
X. Liu, who made some fluid computations on the Tianhe-I machine at National Supercomputing Center, Tianjin, China.

\bibliographystyle{apsrev4-1}
\bibliography{scibib}

\newpage
\begin{figure}[ht!]
 \begin{center}
    \includegraphics[width=110mm]{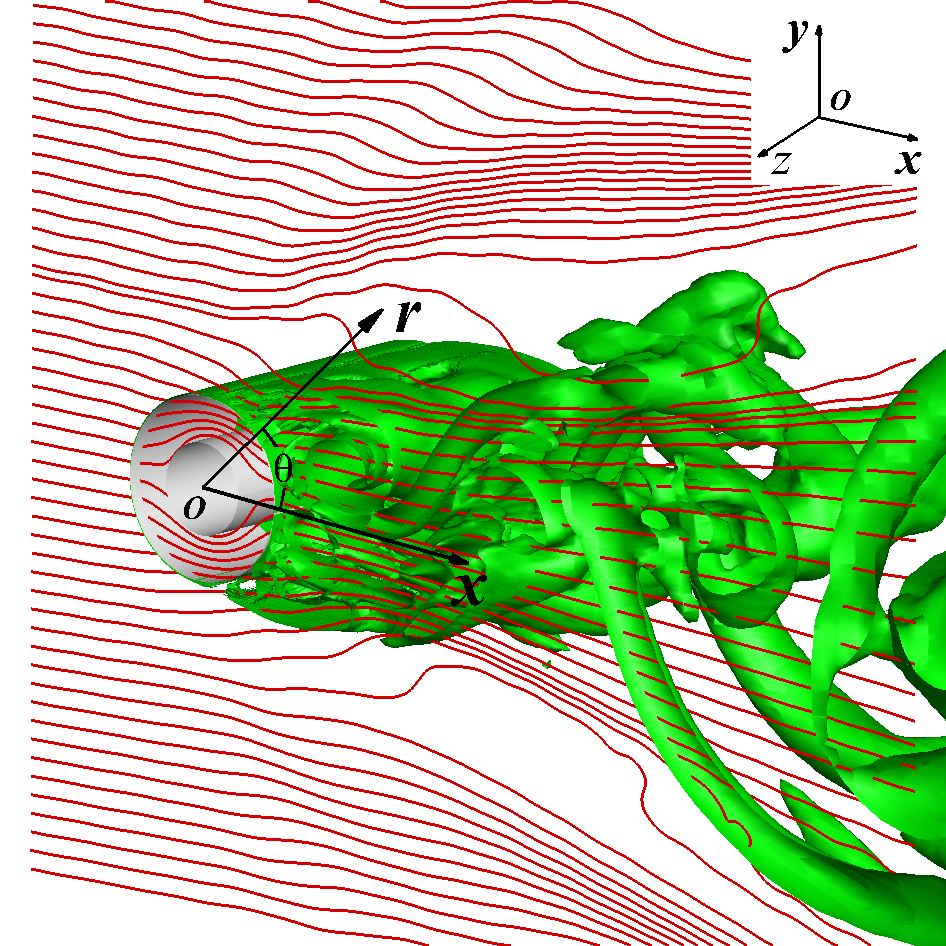}
\end{center}
\caption{ The instantaneous turbulent flow structures (green) and the ray trajectories of the sound waves (red) are calculated using our in-house fluid and acoustic programs.
The flow structures are shown by the $Q$ criterion. Both the free stream and the incident sound wave are directed along the positive $x$-axis.
   }\label{f:turbul}
\end{figure}

\newpage

\begin{figure}[ht!]
\begin{center}
\begin{minipage}{165mm}
\begin{center}
              {\includegraphics[width=160mm]{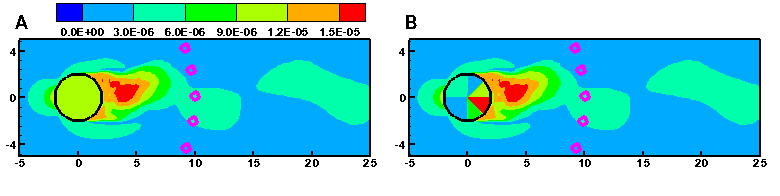}}\\
              {\includegraphics[width=80mm]{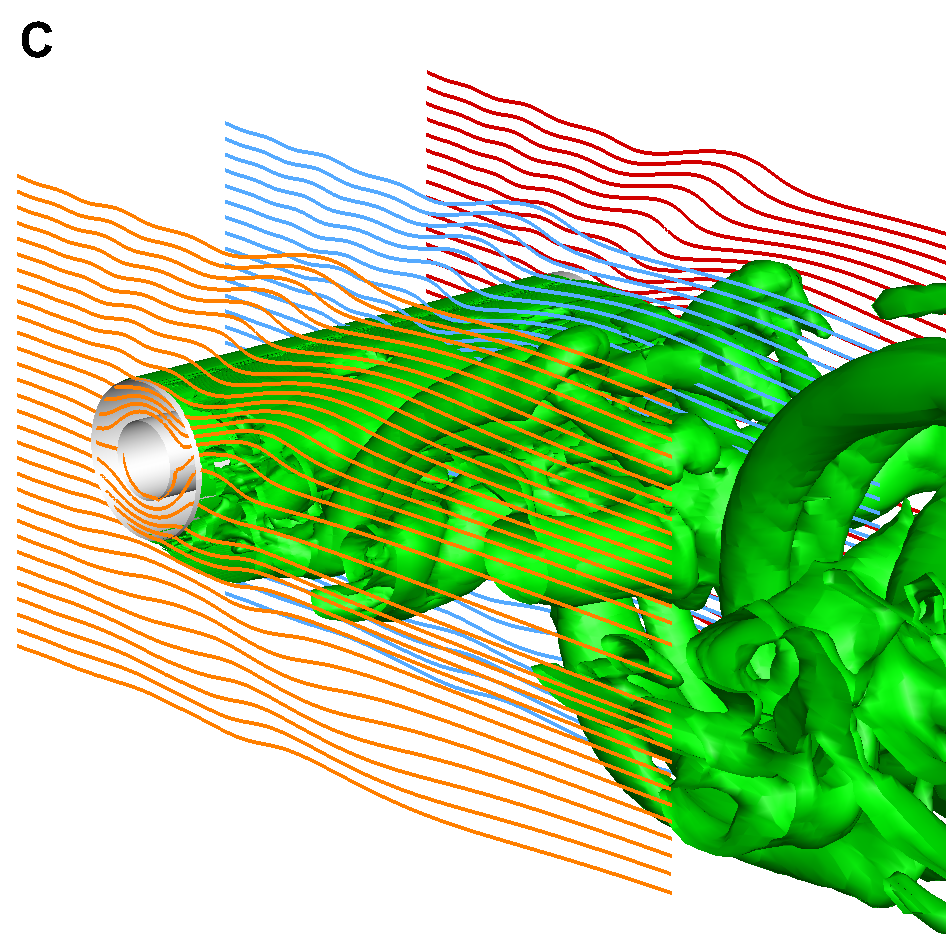}}
      \caption{  \label{f:} The amplitudes of the equivalent sound sources that are driven by the incident plane wave of $k_0=2$.  The purple circles represent the observers that construct the objective function $F_c$. (A) The black circle defines the boundary of a classical cloak designed by transformation acoustics. (B) The black circle defines the boundary of our optimized cloak with inhomogeneous density distributions in the virtual space. (C) The ray trajectories of sound waves at three 2D cross section maps; the optimized cloaking design is used in this simulation; the distortions of ray trajectories in the turbulent fluids are quite small in comparison with the classical cloaking case in Fig. 1.
            }
\end{center}
\end{minipage}
\end{center}
\end{figure}

\newpage

\begin{figure}[ht!]
\begin{center}
\begin{minipage}{165mm}
\begin{center}
              {\includegraphics[width=135mm]{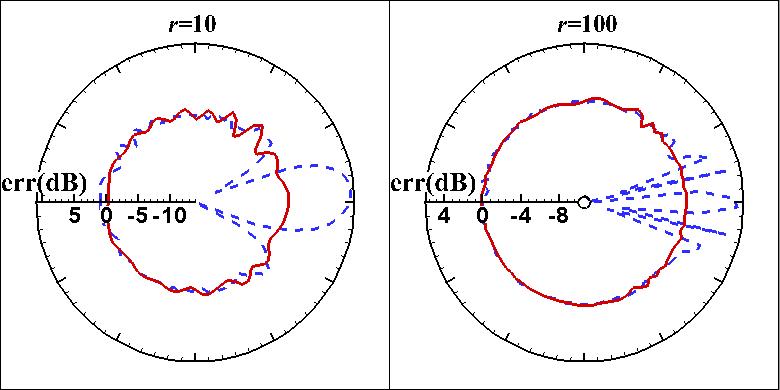}}
      \caption{  \label{f:}  The spatial distributions of $\mathrm{err}$ are shown at $r=10$ and $r=100$, respectively, with respect to $\theta$. Blue lines correspond to the classical cloaking case; red lines correspond to the optimized cloaking case.
       }
\end{center}
\end{minipage}
\end{center}
\end{figure}

\end{document}